\documentclass[twocolumn,journal]{IEEEtran}
\usepackage[T1]{fontenc}
\usepackage[latin9]{inputenc}
\usepackage{xcolor}
\usepackage{pdfcolmk}
\usepackage{array}
\usepackage{calc}
\usepackage{amstext}
\usepackage{graphicx}
\PassOptionsToPackage{normalem}{ulem}
\usepackage{ulem}
\usepackage[unicode=true,
 bookmarks=true,bookmarksnumbered=true,bookmarksopen=true,bookmarksopenlevel=1,
 breaklinks=false,pdfborder={0 0 0},pdfborderstyle={},backref=false,colorlinks=true]
 {hyperref}
\hypersetup{
 pdfauthor={Chenguang Wan},
 pdfpagelayout=OneColumn, pdfnewwindow=true, pdfstartview=XYZ, plainpages=false, citecolor=blue,linkcolor=blue,urlcolor=blue}

\makeatletter

\providecommand{\tabularnewline}{\\}
\providecolor{lyxadded}{rgb}{0,0,1}
\providecolor{lyxdeleted}{rgb}{1,0,0}

\DeclareRobustCommand{\lyxsout}[1]{\ifx\\#1\else\sout{#1}\fi}

\let\oldforeign@language\foreign@language
\DeclareRobustCommand{\foreign@language}[1]{%
  \lowercase{\oldforeign@language{#1}}}

\usepackage[caption=false,font=normalsize,
	labelfont=sf,textfont=sf]{subfig}
\usepackage{cite}

\makeatother

\begin{document}
\title{A robust and fast data management system for machine learning research
of tokamaks}
\author{Chenguang Wan, Zhi Yu, Xiaojuan Liu, Xinghao Wen, Xi Deng, and Jiangang
Li}
\author{\thanks{This work was supported by the National Key R\&D project under Contract
No.Y65GZ10593, the National MCF Energy R\&D Program under Contract
No.2018YFE0304100, and the Comprehensive Research Facility for Fusion
Technology Program of China under Contract No. 2018-000052-73-01-001228.
\emph{(Corresponding authors: Chenguang Wan, Jiangang Li.)}}\thanks{Chenguang Wan, Xinghao Wen, Xi Deng are with the Institute of Plasma
Physics, Chinese Academy of Sciences, Hefei 230031, China, and with
the University of Science and Technology of China, Hefei 230026, China.
(e-mail: \protect\href{mailto:chenguang.wan@ipp.ac.cn}{chenguang.wan@ipp.ac.cn},
\protect\href{mailto:chenguang.wan@ipp.ac.cn}{chenguang.wan@ipp.ac.cn}\protect\href{mailto:wenxh@mail.ustc.edu.cn}{wenxh@mail.ustc.edu.cn},
\protect\href{mailto:xi.deng@ipp.ac.cn}{xi.deng@ipp.ac.cn})}\thanks{Zhi Yu, Xiaojuan Liu, and Jiangang Li are with the Institute of Plasma
Physics, Hefei Institutes of Physical Science, Chinese Academy of
Sciences, Hefei 230031, China (e-mail: \protect\href{mailto:yuzhi@ipp.ac.cn}{yuzhi@ipp.ac.cn},
\protect\href{mailto:lxj@ipp.ac.cn}{lxj@ipp.ac.cn}, \protect\href{mailto:j_li@ipp.ac.cn}{j\_li@ipp.ac.cn}).}}
\markboth{IEEE Transactions on Plasma Science}{Chenguang Wan \MakeLowercase{\emph{et al.}}: A robust and fast data
management system for machine learning research of tokamaks}
\maketitle
\begin{abstract}
In recent years, machine learning (ML) research methods have received
increasing attention in the tokamak community. The conventional database
(i.e., MDSplus for tokamak) of experimental data has been designed
for small group consumption and is mainly aimed at simultaneous visualization
of a small amount of data. The ML data access patterns fundamentally
differ from traditional data access patterns. The typical MDSplus
database is increasingly showing its limitations. We developed a new
data management system suitable for tokamak machine learning research
based on Experimental Advanced Superconducting Tokamak (EAST) data.
The data management system is based on MongoDB and Hierarchical Data
Format version 5 (HDF5). Currently, the entire data management has
more than 3000 channels of data. The system can provide highly reliable
concurrent access. The system includes error correction, MDSplus original
data conversion, and high-performance sequence data output. Further,
some valuable functions are implemented to accelerate ML model training
of fusion, such as bucketing generator, the concatenating buffer,
and distributed sequence generation. This data management system is
more suitable for fusion machine learning model R\&D than MDSplus,
but it can not replace the MDSplus database. The MDSplus database
is still the backend for EAST tokamak data acquisition and storage.
\end{abstract}


\IEEEpeerreviewmaketitle{}

\section{Introduction}

\IEEEPARstart{E}{xperimental} data-driven machine learning (ML) approaches
have been successfully applied to solve various problems in the tokamak
community. These problems include data-driven physic model \cite{Howell2020,Olofsson2013,Moreau2015a,Long2018,DeSilva2020},
disruption prediction \cite{kates-harbeck2019be,Yang_2019,Rea2018,Yokoyama2021b},
magnetic field control \cite{Degrave2022}, surrogate model \cite{clayton2013electron,coccorese1994identification,Bishop1994,Jeon2001,Wang2016a,Joung2020},
experimental data analysis \cite{Samuell2021,Lee2021,Dinklage2008,Churchill2020},
discharge modeling \cite{wan2022,wan2021}, experimental workflow
optimization \cite{Nisan2001,Roughgarden2010,Duris2020,wan2021},
magnetic field reconstruction \cite{Wan2022b}. Generally, the data
access mode for machine learning workflows fundamentally differs from
the conventional access mode for magnetic fusion experiments or simulation
studies. The conventional database (i.e., MDSplus \footnote{MDSplus is a set of software tools for data acquisition and storage
and a methodology for management of complex scientific data. \cite{mongodb}}) of magnetic confinement fusion has been designed for a small group
in the control room, and its primary purpose is to visualize small
amounts of data simultaneously \cite{Anirudh2022}. In significant
contrast, ML data access modes are driven by algorithms that read
and use large amounts of data. Further, data cleaning, normalization,
and generation are critical components of successful fusion ML research.
These fusion ML research key problems are outlined in the Report of
Workshop on Advancing Fusion with Machine Leaning \cite{Humphreys2020}.
In particular, the report supports a new Fusion Data Platform intend
for machine learning research. The platform development includes a
more suitable data management system design for fusion machine learning
R\&D.

The current common tokamak database, MDSplus, does not meet the needs
of the tokamak community for ML research. The data-driven ML approaches
require large amounts of data and have to process long sequences with
different lengths of inputs, whereas typical simulation approaches
\cite{Falchetto2014,Li2022b} do not require such sizable experimental
data inputs. Specifically, in the fusion recurrent neural network
(FRNN) \cite{kates-harbeck2019be}, one of the most famous disruption
prediction works, 8959 shots were used, which would be unimaginable
in the typical simulation code development. Except for the requirements
of large amounts of data, fusion ML research requires high Input/Output
(I/O) data read/write, MapReduce \cite{Dean2008} metadata calculation,
data alignment, data conversion, sequence partitioning, error correction,
sequence concatenating, and distributed operation support. MapReduce
is a programming model and an associated implementation for processing
big data sets with a parallel, distributed algorithm on a cluster.
These operations are not supported by the existing MDSplus database
of tokamak \cite{wang2018studyof}.

To meet the new requirements of the experimental data-driven fusion
ML research, we propose a new data management system that combines
MongoDB and HDF5 \cite{Folk2011} files and develops some backend
engines more suitable for tokamak fusion ML research. MongoDB is suitable
for data retrieval and preprocessing, while the entry of MongoDB is
limited to 64 megabytes (MB). The Hierarchical Data Format version
5 (HDF5) is designed to store and organize large amounts of data,
and it is also suitable for high I/O data access since HDF5 files
are file-level data management. The file-level data management features
also cause HDF5 files are hard to retrieve and preprocess. We design
a hybrid database that combines the advantages of MongoDB and HDF5.

The rest of this paper consists of four parts. Section \ref{sec:System-design}
details the data management system design, including low-level data
organization and high-level operation engines. Section \ref{sec:Performance}
shows the performance comparison between the conventional MDSplus
database and this work. Finally, a brief discussion and conclusion
are given in Section \ref{sec:Discussion}

\section{System design \label{sec:System-design}}

\subsection{Architecture}

\begin{figure*}[tbh]
\begin{centering}
\includegraphics[height=0.75\textheight]{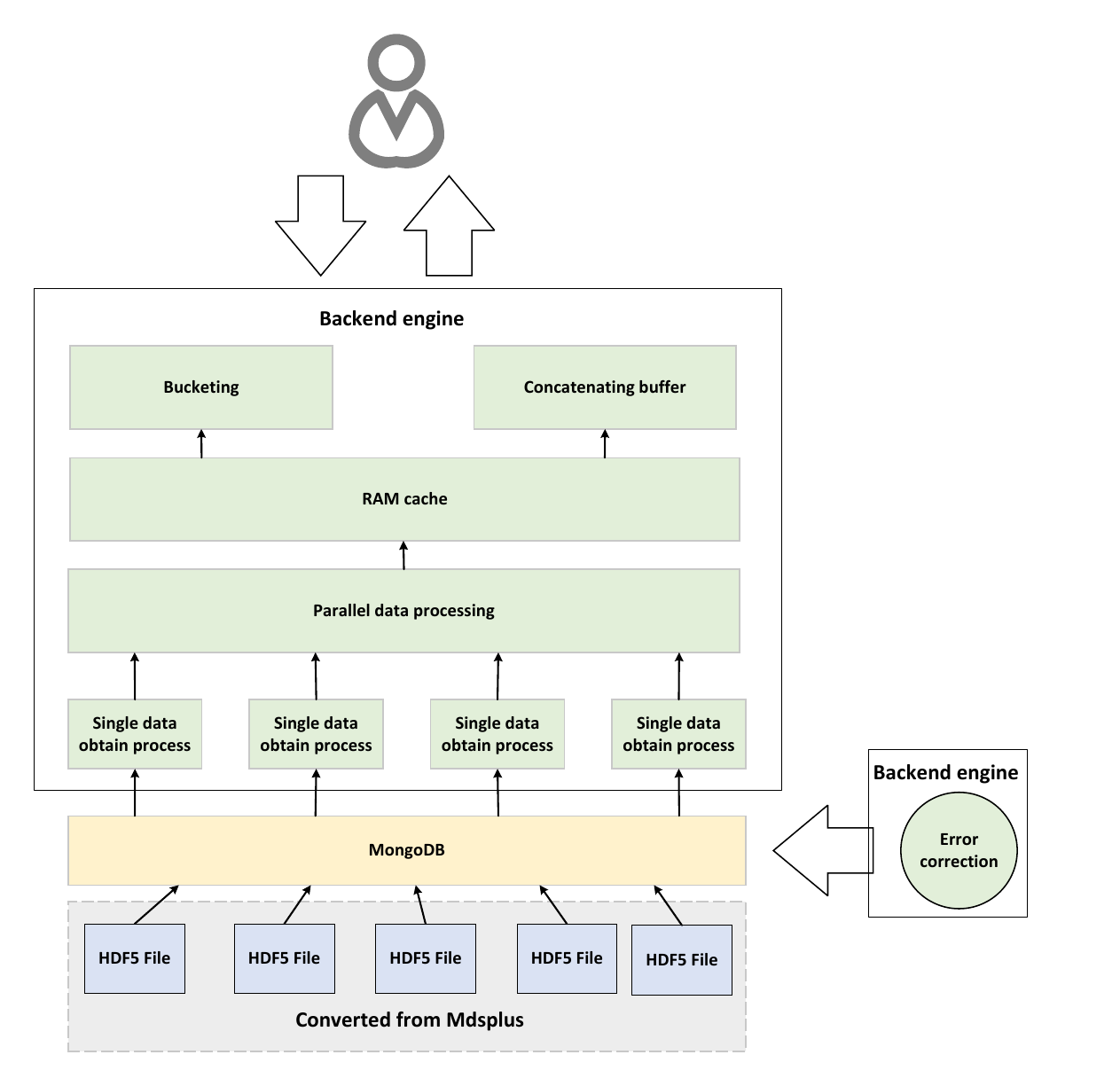}
\par\end{centering}
\caption{The data management system architecture. The architecture is divided
into three parts, the HDF5 format low-level original data (blue boxes),
the metadata managed using MongoDB (yellow box), and the high-level
data operation engines (light green boxes).\label{fig:workflow}}
\end{figure*}

The system is designed based on a combination of MongoDB and HDF5.
The data management system is divided into three parts the low-level
original data management, the metadata management, and the operational
engines. Fig. \ref{fig:workflow} shows our data management system
workflow., The yellow box is metadata management, the blue boxes are
low-level original data , and the light green boxes are data operational
engines. The main operations are listed as follows:
\begin{enumerate}
\item The MDSplus tree data is converted shot-by-shot to HDF5 files.
\item Correct the original HDF5 files with the error correction engine.
\item MongoDB contains metadata about the HDF5 files, such as their location,
node statistical properties, etc.
\item The single process is used to read an HDF5 file.
\item Parallel processing is called by bucketing generator, concatenating
buffer, and calculating the global metadata.
\item Set bucketing generator (explained further in Section \ref{subsec:Engine-design})
boundaries.
\item Get aligned sequences from the bucketing generator.
\item Set concatenating buffer time step $\delta$.
\item Get concatenated sequences from the buffer.
\end{enumerate}

\subsection{Database design}

The MDSplus database is designed as a tree structure to store the
complex tokamak experimental data. The MDSplus database is unsuitable
for fusion ML data accessing and processing. However, recently, experimental
data-driven ML research in tokamaks is becoming increasingly popular,
and these methods need high robustness, parallelism, and high I/O
support of the database or the data management system and also require
the database to have the ability to compute global metadata in the
distributed MapReduce way.

Our architecture is designed based on MongoDB and HDF5 to meet the
new requirements of fusion ML research on tokamaks. MongoDB is a source-available
cross-platform document-oriented database program. Classified as a
NoSQL database \footnote{The NoSQL (aka ``not only structured query language'') database
is a non-tabular database and stores data differently than relational
tables} program, MongoDB uses JSON-like \footnote{JSON is an open standard file format that uses readable text to store
data objects consisting of attribute-value pairs and arrays.} documents with optional schemas and supports high I/O. MongoDB also
with the ability for high-performance data accessing. Although MongoDB
has lots of advantages, on the one hand, even with the latest MongoDB,
the maximum entry size is only 64 megabytes (MB). One tokamak experiment
usually produces over one gigabyte (GB) of data. On the other hand,
if each entry had a considerable size, the retrieval performance and
robustness would be bad. We use HDF5 files saved as low-level source
data storage to solve this problem. The HDF5 is an open-source file
format that supports large, complex, heterogeneous data. HDF5 uses
a \textquotedbl file directory\textquotedbl{} structure that allows
users to organize data within the file in many different structured
ways, as you might do with files on your computer. The HDF5 format
also allows for metadata embedding, making it self-describing. The
HDF5 is robust and supports high I/O since it is only a file stored
on disk, not a server layer required. Our approach is a hybrid architecture
of MongoDB and HDF5. The approach uses MongoDB as the data indexing
layer and stores metadata in the MongoDB database. As shown in Table
\ref{tab:MongoDB-data-structure}, the metadata includes the discharge
duration time, every signal mean, variance, existence flag, correction
flag, etc. The statistical summary is used for data filtering, and
the correction flag is used to mark whether the corresponding HDF5
node is corrected. The original MDSplus data is converted and saved
on original HDF5 files on the cluster disk.

\begin{table}[tbh]
\caption{The MongoDB entry structure \label{tab:MongoDB-data-structure}. <node>
is a general name for EAST diagnostic signal. It can be replaced by
WMHD, Ne, etc.}

\centering{}%
\begin{tabular}{ccc}
\hline 
Key & type & meaning\tabularnewline
\hline 
shot & int & Tokamak experiment shot number\tabularnewline
file\_location & string & The corresponding HDF5 file location\tabularnewline
discharge\_time & float & Discharge duration time\tabularnewline
<node>\_existence & bool & Signal existence of this shot\tabularnewline
<node>\_corrected & bool & Signal correction flag\tabularnewline
<node>\_mean & float & Signal mean\tabularnewline
<node>\_stDev & float & Signal standard deviation\tabularnewline
<node>\_start & float & Node start time\tabularnewline
<node>\_end & float & Node end time\tabularnewline
\hline 
\end{tabular}
\end{table}

\subsection{Engine design \label{subsec:Engine-design}}

We have developed some operation engines for this data management
operation because this hybrid data management system is not easy to
operate directly. The engines have six main components: Single data
obtaining process, parallel data processing, bucketing generator,
concatenating buffer, error correction and data conversion. The single
data obtaining process is used to read a shot of data from the hybrid
data management system, the different signals data will be aligned
with an identical time axis. Users define the time axis start and
end, and the default value \textquotedbl start\textquotedbl{} is
equal to zero, and \textquotedbl end\textquotedbl{} is equal to the
discharge end time. The parallel data obtain process calls the single
data obtaining process to get multiple shots data. Bucketing generator
sets some parameters that fit the entire discharge research sequence
generation of fusion ML research and caches some data in memory to
solve the different speeds from data generation and ML model training.
Concatenating buffer concatenates sequences from different shots to
generate the aligned subsequences. The error correction is used to
check for NaN (invalid value) and Inf (infinity value) read from the
original data. In this step, if the input data has NaN or Inf will
be replaced by a linear interpolation value and $3.2\text{\ensuremath{\times10^{32}}}$(it
is not the maximum value of the float 32 type, but it is large enough.
And it can still be calculated without overflowing), respectively.
If the node is corrected, the corrected flag will be set as true.
The data conversion is used to convert the raw MDSplus data to HDF5
files.

The mini-batching gradient descent \cite{dean2012large,Huang2013}
is a general technique that helps enhance GPU performance \cite{Chetlur2014}
and accelerates the training convergence of ML models. The loss gradients
are computed for several examples in parallel and then averaged. The
ML model training of the tokamak's experimental data is difficult
to use the mini-batching since the tokamak experiment is all time
sequences and has different sequence lengths because of the different
duration of each tokamak experiment. For the ML approach to work efficiently,
the ML model training for forward and backward passes of each gradient
computation must be equal for all the examples computed in parallel.
This is not possible if different training examples have different
lengths. Therefore, there are three options for ML model training
of experimental data of tokamaks. 1. Using time slicing techniques,
ML models input time slices instead of sequences. 2. Time sequence
padding techniques, padding sequences with specific values to the
same length. 3. Time sequence concatenating. Option 1 is straightforward
to implement, so the present work does not give the corresponding
interface. A method called a \textquotedbl bucket generator\textquotedbl{}
was developed in the engine to satisfy the requirements of Option
2. For Option 3, we developed an approach based on a sequence concatenating
buffer to meet this requirement.

\subsection{Bucketing generator design}

Tokamak experiment durations vary from experiment to experiment and
generate sequences of different lengths. A mini-batch must train on
data of the same length, and the short sequences are padded. Therefore,
if the sequences are shuffled, a mini-batch could have wildly different
maximum and minimum sequence lengths and be action on a lot of padded
data. This increases training time and memory usage.

Bucketing is a partitioning algorithm that is used to speed up the
training time of a long sequence dataset. We assume there is a set
of sequence $S=\left\{ s_{1},s_{2},\ldots,s_{n}\right\} $ is our
train set. $l_{i}=\left|s_{i}\right|$ is the length of sequence $s_{i}$.
And we use a mini-batch approach to train a ML model for that train
set. Each GPU processes a mini-batch sequences in a synchronized parallel
manner, so a mini-batch $I_{\text{batch}}=\left\{ s_{1},s_{2},\ldots,s_{k}\right\} $
cost time is proportional to $O\left(\text{max}_{i\in1,\ldots,k}l_{i}\right)$.
Therefore, the processing time of the entire train set is expressed
as:

\begin{equation}
\text{T}(S)=O\left(n/k\text{\ensuremath{\times}}\text{max}_{i\in1,\ldots,k}l_{i}\right).\label{eq:1-1}
\end{equation}

If the sequences of the train set were shuffled randomly before mini-batch
generation, mini-batch's minimum and maximum sequence lengths would
be very different. As a result, the GPU would do useless work for
processing the meaningless padding tails of shorter sequences. Additionally,
the long sequences with a bit bigger batch size would reach GPU memory
capacity limit.Specifically, using the same mini-batch size for long
sequences input as for short sequences input takes up more GPU memory
than expected. We develop a customized bucketing generator backend
to optimize the batch training to overcome this flaw and reduce training
time. The entire sequence data set is partitioned into $B$ buckets
by the lengths, where each bucket contains data with a similar sequence
length. Let $S_{i}=\left\{ s_{j_{1}},s_{j_{2}},\ldots s_{j_{k_{i}}}\right\} $.
For every bucket, we perform the mini-batch training with different
batch sizes. The processing time of the whole set is expressed as:

\begin{equation}
T(S)=\sum_{i=1}^{B}O(\text{T}(S_{i})).\label{eq:}
\end{equation}

The sequences within every bucket are shuffled randomly. And then,
the sequences are generated batch sequences batch-by-batch. To train
batchwise with a batch size $M$, we need M independent shot discharge
sequences of the same bucket to feed to the GPU. Zeros pad the different
length discharge sequences to the same length. We do this by using
M processes to read sequence data in parallel. The M sequences are
fed to a buffer first to solve the problem of GPU and CPU speed mismatch
since data from HDF5 files are read through a CPU.

\begin{figure}[tbh]
\begin{centering}
\includegraphics[width=1\columnwidth]{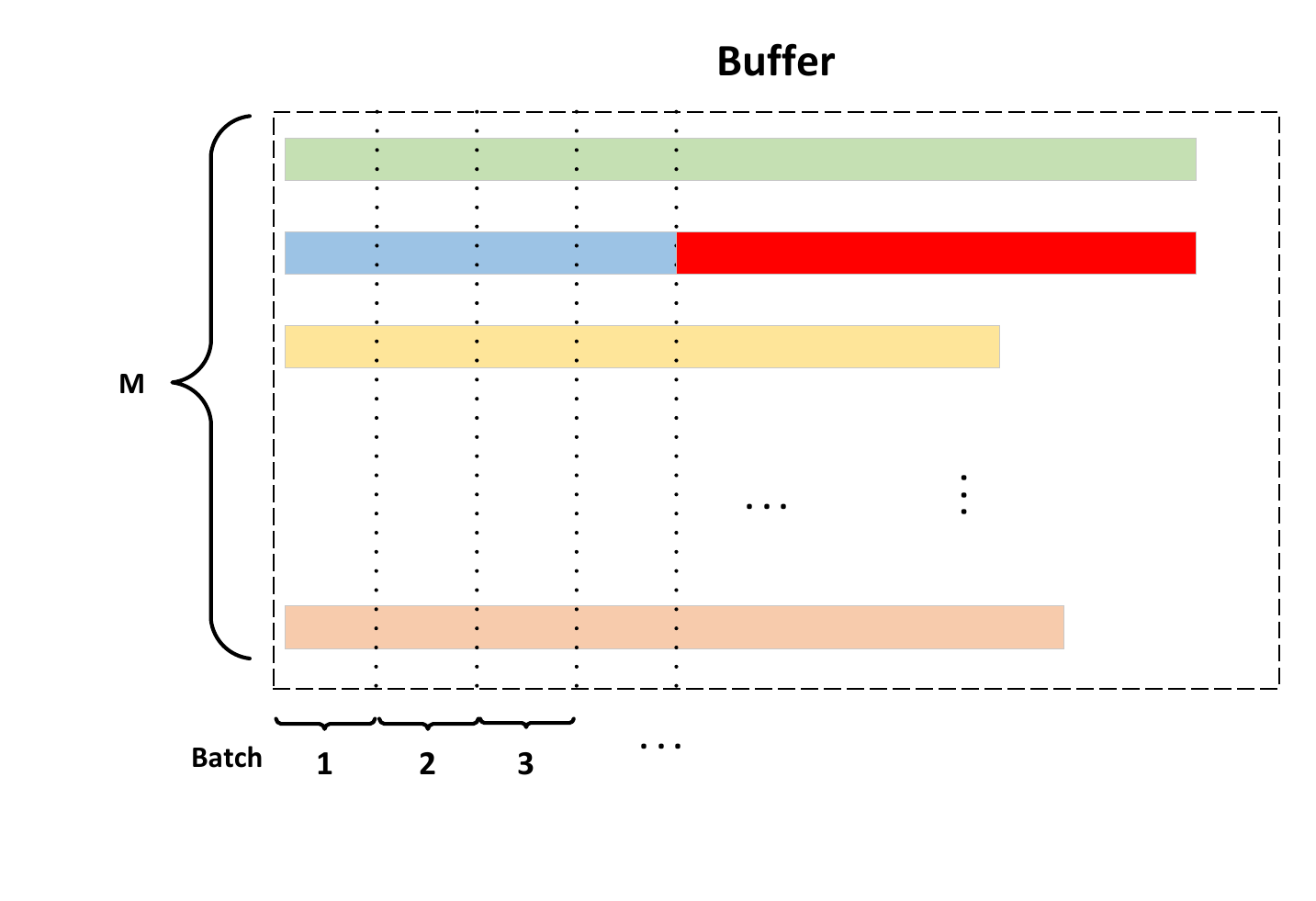}
\par\end{centering}
\caption{\label{fig:Buffer}Sketch of the concatenating buffer engine design.
The figure illustrates how batchwise data generate from the buffer
with batch size M. The different horizontal bars with different colors
represent different shots . A color change in a given row means that
a new shot starts. At every time step, the leftmost chunk is cut from
the buffer and output. Generally, the concatenating buffer can be
regarded as a window moving on the shot sequence set.}
\end{figure}

\subsection{Concatenating buffer design}

Another way to handle variable-length time sequence data is concatenating,
which is also supported by our data management system engine. Fig.
\ref{fig:Buffer} shows the concatenating buffer design. In Fig. \ref{fig:Buffer},
a concatenating buffer to output M equal length sequences to feed
to the ML model. The sequences come from different shots. Firstly,
we need to set a time step $\delta$, before using the concatenating
buffer. $\delta$ is the minimum time-dependent length. For example,
the time-dependent length in disruption research is always set to
$\sim0.1$ s \cite{kates-harbeck2019be}. In the buffer, shot sequences
are cut into multiples of $\delta$ chunks at the beginning. Whenever
a shot finishes processing (e.g., the blue shot in Fig. \ref{fig:Buffer}),
a new shot (red) is loaded. The chunks that are successive in the
shot must also be successively generated mini-batches such that the
ML model's internal state can be passed correctly. In the concatenating
buffer, shot data is not read all at once but is read asynchronously
by calling the data reading engine. In practice, there is a data maintenance
process that ensures that there is always enough data in the buffer.
The concatenating buffer can be regarded as a window that moves through
the discharge sequence group. The buffer allows the training of shots
of different lengths in batches.

The concatenating buffer data input and the bucketing generator data
input are suitable for different ML models training. For example,
suppose the entire tokamak discharge modeling is the target of ML
research \cite{wan2021}. In that case, the bucketing method is a
good choice. The concatenating buffer is a general option if the ML
model is to study local time dependencies (like disruption prediction).
On the other hand, if the ML model does not require time-dependent,
the easily slicing approach is more common.

\subsection{Data conversion and error correction}

The dataset is selected from the EAST tokamak MDSplus database original
data. Since the MDSplus database cannot support high I/O data accessing,
we developed a data conversion engine (see Fig. \ref{fig:workflow})
to convert MDSplus original data to HDF5 files shot-by-shot directly,
and one shot is one HDF5 file. And then, we use the error correction
engine to correct the original HDF5 files to get corrected HDF5 files.
The high-level engine operates corrected HDF5 files by default (changeable),
but we also provide APIs to support custom data error correction or
to set it to not correct.

\section{Performance analysis \label{sec:Performance}}

In this section, we compare in detail the performance of our system
with the convention database in tokamaks. We compare the bucketing
with conventional shuffling and do not compare the concatenating buffer.
The concatenating buffer is a solution-suit tool for specific machine
learning model training and does not have a counterpart in the conventional
fusion database.

\subsection{Comparing hybrid data management and MDSplus}

Highly concurrent data accessing directly through the server in EAST's
real MDSplus database are difficult to achieve. With so many people
working on the EAST's MDSplus database, we can't perform high-risk
operations on it. So we built a simple MDSplus database for data reading
efficiency comparison. The demonstrative database contains 100 signal
data from the EAST original shot in range \#74000-76000. We compared
the time to read 100 signal data in shot range \#74000-76000 from
MDSplus and our data management system. The simple MDSplus and our
system were deployed in the same cluster to minimize interference
from different environments. Although our model has high concurrency
support, we used four processes in both systems for data reading in
this test since the MDSplus does not have high concurrency. This means
that the data access to our system is faster than reported. The results
are contained in Table \ref{tab:Comparison}. We also listed some
common function comparisons in ML, simulation, and experiment research
in Table \ref{tab:Comparison}.

\begin{table}[tbh]
\caption{Comparison of our hybrid data management and MDSplus. \label{tab:Comparison}}

\centering{}%
\begin{tabular}{cc>{\centering}p{0.25\columnwidth}}
\hline 
Function & MDSplus & Our hybrid data management system\tabularnewline
\hline 
Demonstrative reading time & $\sim4289$ s & $\sim167$ s\tabularnewline
Bucketing support & False & True\tabularnewline
Concatenating support & False & True\tabularnewline
High IO concurrency & False & True\tabularnewline
ML Algorithm Support & Normal & Good\tabularnewline
Visualization support & True & False\tabularnewline
Number of plug-ins & Many & None\tabularnewline
Diagnostic raw data interface & True & False\tabularnewline
\hline 
\end{tabular}
\end{table}

Table \ref{tab:Comparison} shows that our model has an advantage
in ML-related operational support. MDSplus, on the other hand, has
the advantage of supporting mainly traditional simulation studies,
and it can be directly interfaced with fusion diagnostic systems,
which is something our system cannot do at the moment. Our data management
is good for fusion ML research but can not substitute the MDSplus
database.

\subsection{Bucketing performance}

\begin{figure}[tbh]
\centering{}\includegraphics[width=1\columnwidth]{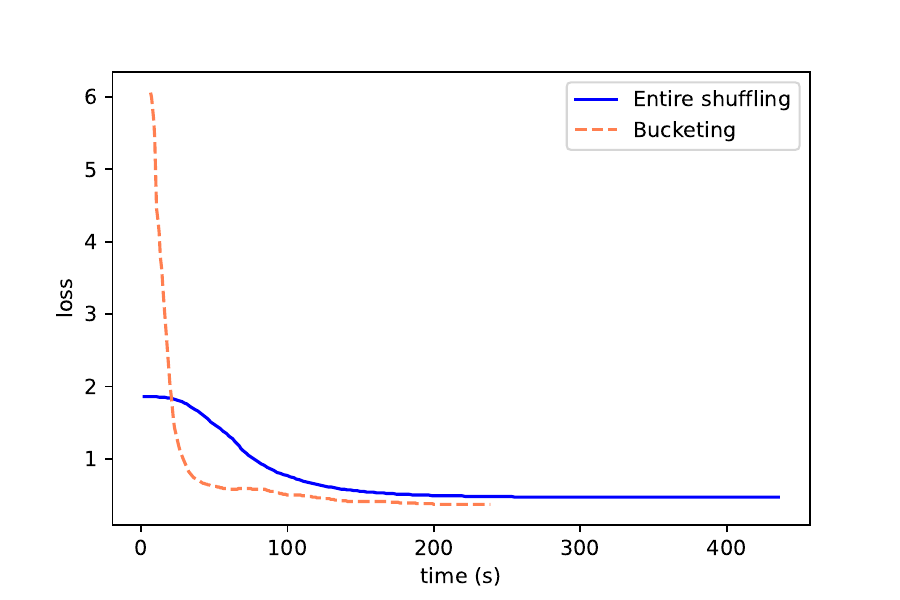}\caption{\label{fig:Training_time}Training time comparison. The ML model using
bucketing algorithm training converges in about 50s, while the ML
model using shuffling algorithm training converges in about 110s.
This means that the bucketing algorithm is twice as efficient as the
common shuffling algorithm in the current task.}
\end{figure}

We tested the ML model's training time in 300 shots sampled from the
EAST 2020-2020 campaigns. Fig. \ref{fig:Training_time} compares the
bucketing approach and the entire dataset shuffling approach. In this
example, we trained a simple single layer long-short term memory (LSTM)
neural network with the same learning rate, optimizer, weights initializer,
bias initializer, loss function, etc. The input parameters come from
the experiment data-driven discharge modeling \cite{wan2021}, and
the output parameter is stored energy $W_{mhd}$. The bucketing is
more efficient than conventional random shuffling in the ML model
training. The blue line of Fig. \ref{fig:Training_time} is random
shuffling and this approach converges more slowly than the bucketing.
We provide a bucketing backend engine that is able to speed up the
training of the ML model.

\section{Discussion and Conclusion \label{sec:Discussion}}

As the fusion community is increasing interest in data-driven ML research,
the traditional MDSplus database is increasingly showing its limitations
in ML research. To fully harness the transformative potential that
ML might provide in many fusion energy-related fields, these flaws
must be remedied. The idea of Fused Data Platforms (FDP)\cite{Humphreys2020}
for ML research is gaining more and more attention. In the present
work, A new data management system suitable for fusion ML model R\&D
of EAST tokamak has been designed and developed to primary achieve
the idea. The system not only accommodates algorithm-driven, high
I/O tokamak data access but also provides a series of advanced interfaces
for more efficient machine learning data generation. The system has
the following main functions: data conversion from MDSplus original
data, error correction, concatenating buffer sequence generation,
bucketing sequence generation, and data statistics. All functions
have been developed and have been tested on real fusion ML work about
the last closed magnetic surface reconstruction \cite{Wan2022b}.

In future work, the error correction engine will be upgraded by analyzing
each tokamak diagnosis and will process the values exceeding the corresponding
diagnostic limitation. An automatic data cleaning will be developed
to automatically filter out outlier experiments and error experiments.
Further, we intend to focus on building a platform to provide an integrated
environment for machine learning and data exploration studies supported
by a common interface. Finally, We will test our system in several
tokamak databases and develop compatible visualization tools to accelerate
data-driven ML research in the fusion community.

\appendices{}

\section*{Acknowlegment}

The authors would like to thank all the members of EAST Team for providing
such a large quantity of past experimental data. The authors sincerely
thank Prof. Qiping Yuan, Dr. Ruirui Zhang, and Prof. Jinping Qian
for explaining the experimental data.

This work was supported by the National Key R\&D project under Contract
No.Y65GZ10593, the National MCF Energy R\&D Program under Contract
No.2018YFE0304100, and the Comprehensive Research Facility for Fusion
Technology Program of China under Contract No. 2018-000052-73-01-001228.

\bibliographystyle{IEEEtran}
\bibliography{IEEEabrv,library}
\end{document}